\documentstyle[aps,prb,psfig,floats]{revtex}

\begin{document}
\draft
\title{\vspace*{-5mm}\hfill
{\normalsize submitted to Phys.~Rev.~B.}
       \vspace{5mm}\\
Ab-{}initio study of the anomalies in the\\ 
He atom scattering spectra of H/Mo\,(110) and H/W\,(110)}

\author{Bernd Kohler, Paolo Ruggerone, and Matthias Scheffler}
\address{
Fritz-Haber-Institut der Max-Planck-Gesellschaft,\\
Faradayweg 4-6, D-14\,195 Berlin-Dahlem, Germany}
\date{\today}
\maketitle
\begin{abstract}
Helium atom scattering (HAS) studies of the H-covered 
Mo\,(110) and W\,(110) surfaces reveal a twofold anomaly in the 
respective dispersion curves. 
In order to explain this unusual behavior we performed  
density functional theory calculations of the atomic and 
electronic structure, 
the vibrational properties, and the spectrum of electron-{}hole 
excitations of those surfaces.
Our work provides evidence for hydrogen adsorption 
induced Fermi surface nesting. 
The respective nesting vectors are in excellent agreement with
the HAS data and recent angle resolved photoemission experiments 
of the H-{}covered alloy system Mo$_{0.95}$Re$_{0.05}$\,(110). 
Also, we investigated the electron-{}phonon coupling and discovered 
that the Rayleigh phonon frequency is lowered for those critical 
wave vectors. 
Moreover, the smaller indentation in the HAS spectra can 
be clearly identified as a Kohn anomaly. 
Based on our results for the susceptibility and the recently 
improved understanding of the He scattering mechanism we argue 
that the larger anomalous dip is due to a direct interaction of 
the He atoms with electron-{}hole excitations at the 
Fermi level.
\end{abstract}
\narrowtext
\section{Introduction}\label{SIntro}
\begin{figure}
\begin{center}
\hspace{.1mm}
\psfig{file=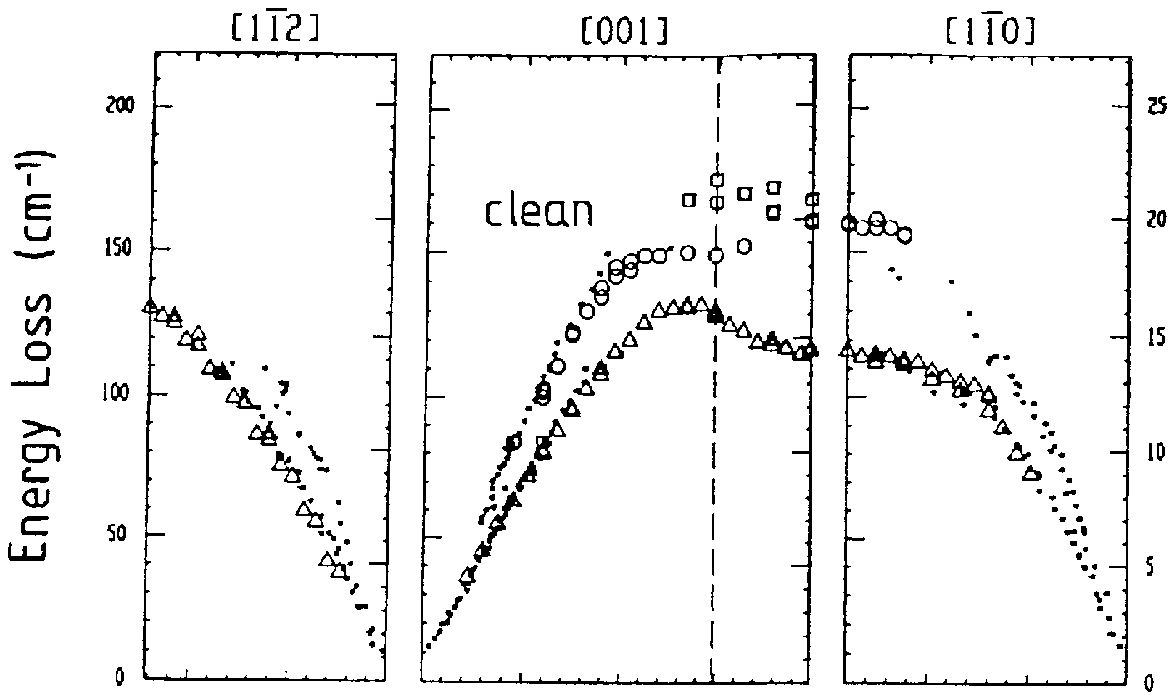,width=8.5cm}
\psfig{file=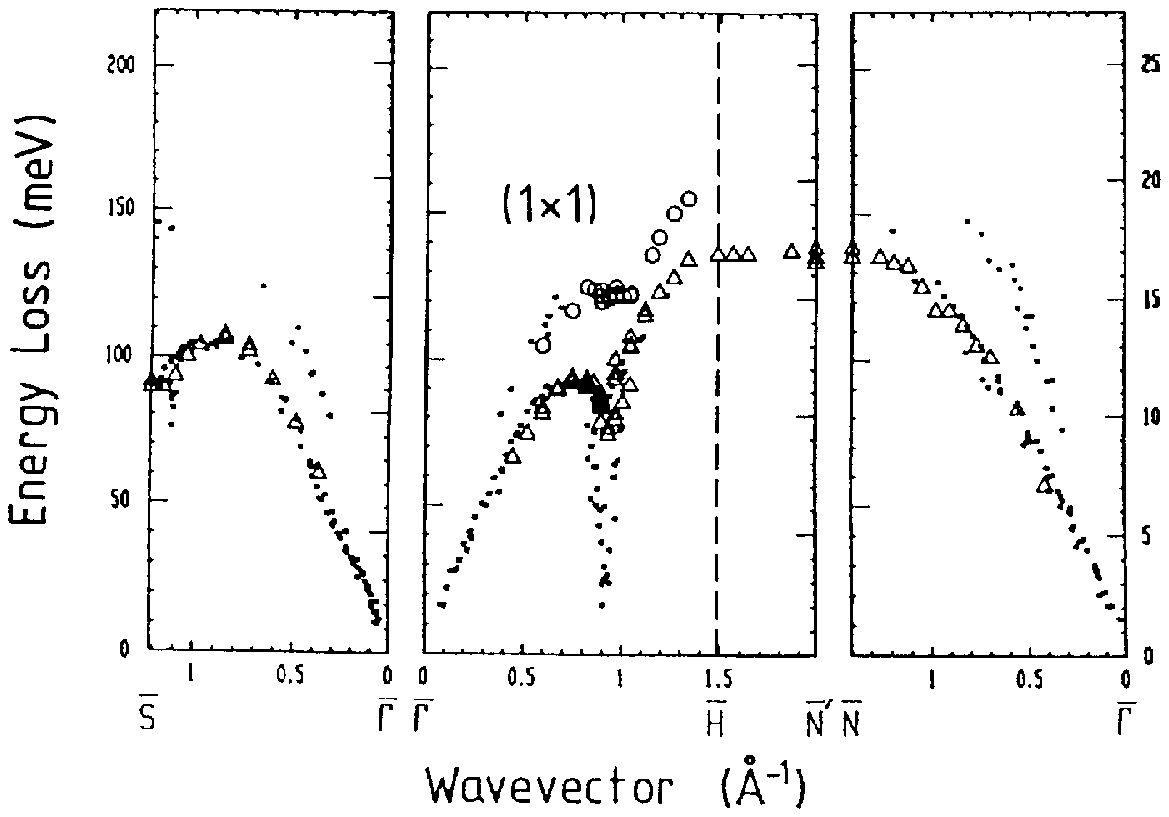,width=8.5cm}
\end{center}
\caption{
HREELS spectra of the clean W(110) surface (upper panel) and 
its H-{}covered $(1\times 1)$ phase (lower panel) from 
Ref.~\protect{\onlinecite{bald95a}}: 
Shown are the dispersion of the Rayleigh 
wave (triangles) and the longitudinally polarized surface phonons 
of the first (circles) and second (squares) layer. 
The dots represent the results of Hulpke and L\"udecke's HAS 
measurements~\protect{\cite{hulp92a,lued94}}.}
\label{FAnom}
\end{figure}
\begin{figure}
\begin{center}
\hspace{.1mm}
\psfig{file=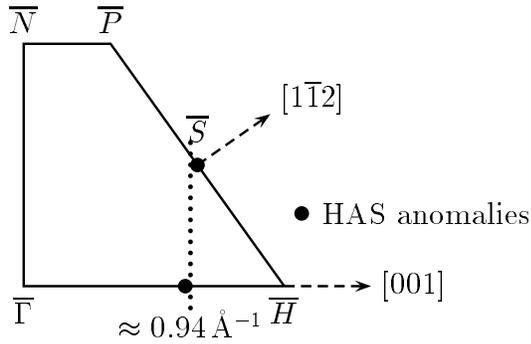,width=7cm}
\vspace{2mm}
\end{center}
\caption{
Position of the HAS anomalies for the H/W(110) adsorbat system 
within the SBZ. 
The dotted line indicates the form of the anomaly apart from 
the symmetry directions \protect{$\overline{\Gamma H}$} 
and \protect{$\overline{\Gamma S}$} as suggested by Hulpke 
and L\"udecke~\protect{\cite{hulp92b}}.} 
\label{FLocus}
\end{figure}

The interest in the (110) surfaces of Mo and W has been fostered 
in recent years by the discovery of deep and extremely sharp 
indentations in the energy loss spectra of He atom scattering at 
H/Mo\,(110) and H/W\,(110)~\cite{hulp92a,lued94}. 
As depicted in Fig.~\ref{FAnom} those anomalies are seen at an 
incommensurate wavevector, $\overline Q^{\rm c,1}_{\rm exp}$, 
along the $[001]$ direction (${\overline{\Gamma H}}$) and additionally 
at the commensurate wavevector 
$\overline Q^{\rm c,2}_{\rm exp} = \overline S$ at the boundary of the surface 
Brillouin zone (SBZ).
At those points two simultaneous anomalies develop out of 
the ordinary Rayleigh mode. 
One, $\omega_1$, is extremely deep, and is only seen by helium atom  
scattering (HAS)~\cite{hulp92a,hulp92b,hulp93a,hulp93b,lued94}. 
The other, $\omega_2$, is instead modest, and is 
observed by both HAS 
and high resolution electron energy loss spectroscopy 
(HREELS)~\cite{bald95a,bald94a,bald94b}. 
Since the $[001]$ components of both critical wavevectors 
$\overline Q^{\rm c,1}$ and $\overline Q^{\rm c,2}$ are approximately 
the same it was suggested that the anomaly runs parallel to 
$\overline{\Gamma N}$ through the SBZ as illustrated 
in Fig.~\ref{FLocus}. 

\begin{figure}
\begin{center}
\hspace{.1mm}
\psfig{file=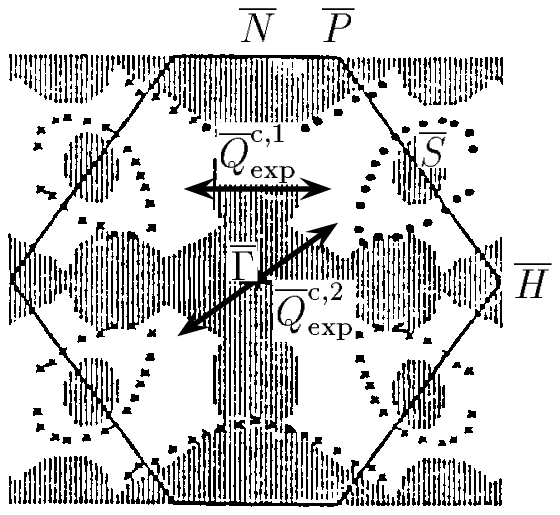,width=7cm}
\vspace{2mm}
\end{center}
\caption{ARP-{}measured Fermi surface of the H-{}covered W\,(110) 
surface~\protect{\cite{gayl89}}. 
The shaded areas are the (110) projection of the bulk Fermi surface. 
Surface states at the Fermi level are marked by dots and repeated 
in the other equivalent sections by crosses. 
The arrows represent the HAS critical wavevectors 
\protect{$\overline Q^{\rm c,1}_{\rm exp}$}
 and \protect{$\overline Q^{\rm c,2}_{\rm exp}$}.}
\label{FFermiExp}
\end{figure}
Various models have been brought forward in order to account for the 
unusual vibrational properties of H-{}covered Mo(110) and W(110):
For instance, a pronounced but less sharp softening of surface phonons 
is also seen on the (001) surfaces of W~\cite{erns92a} and 
Mo~\cite{hulp89}. 
There, the effect is caused by marked nesting properties 
of the Fermi surface~\cite{wang87,wang88a,smit90,chun92}. 
However, angular resolved photoemission (ARP) 
experiments of H/W\,(110) and H/Mo\,(110) 
gave no evidence for nesting vectors 
comparable to the HAS determined critical wave 
vectors~\cite{gayl89,jeon89a,jeon89b} 
(see Fig.~\ref{FFermiExp}). 
Thus, for the phonon anomalies of those two systems 
there seems to be no connection between the electronic structure 
and the vibrational properties.  
Moreover, even if such a relation existed it would be unclear whether the 
phenomena has to be interpreted as a Kohn anomaly or as the phason 
and amplitudon modes associated with the occurance of a 
charge-{}density wave. 
In fact, the experimental finding of two modes at $\overline S$ as  well 
as some theoretical arguments rule out the phason-{}amplitudon 
idea~\cite{kohl96b}. 
Furthermore, any model which links the phonon anomalies to the 
motion of the hydrogen atoms~\cite{bald94a} has to be ruled out 
because the HAS spectra remain practically unchanged when deuterium 
is adsorbed instead of hydrogen~\cite{hulp92a,lued94}. 

Still unexplained, but probably not directly correlated to the anomalies, 
is the peculiar behavior of hydrogen vibrations observed 
in HREELS experiments by Balden {\em et al.}~\cite{bald94a,bald94b}. 
At a H-{}coverage of one mono-{}layer the adsorbate modes parallel 
to the surface form a continuum interpreted a liquid-{}like phase 
of the adsorbate. 

\begin{figure}
\begin{center}
\hspace{.1mm}
\psfig{file=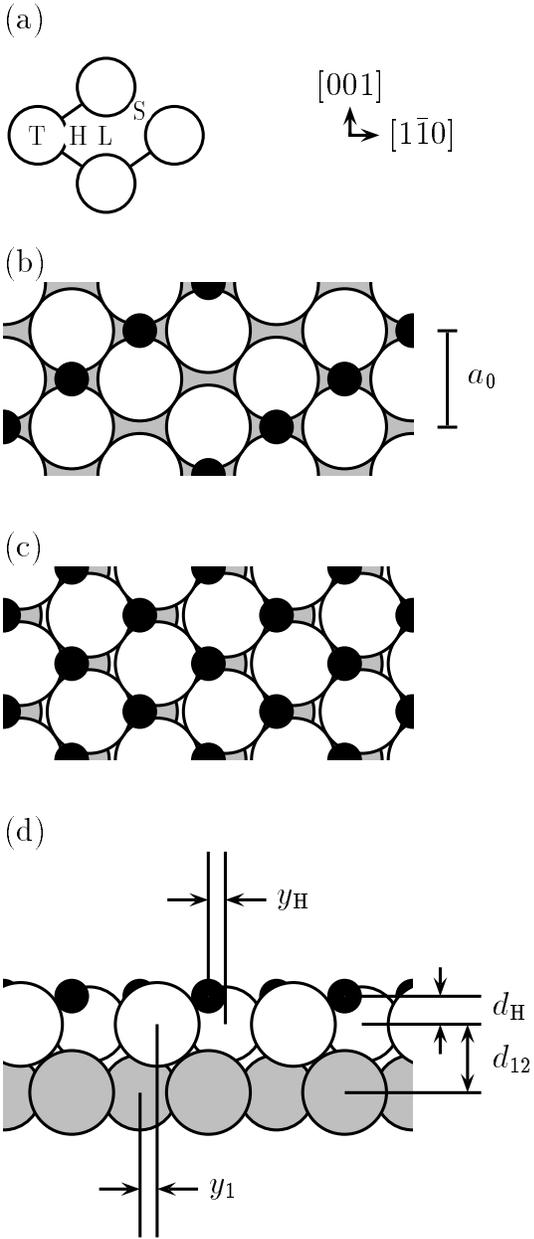,width=7cm}
\vspace{2mm}
\end{center}
\caption{(a) Adsorbate positions within the 
$(1\times 1)$ unit cell of the W\,(110) and Mo\,(110) surfaces: 
Shown are the long-bridge (L), short-bridge (S), hollow (H), 
and on-top (T) site. 
(b-c) Structure of the H-{}covered W\,(110) surface as 
suggested by Chung {\em et al.} \protect{\cite{chun86}}: 
(b) low H-{}coverage $\Theta< 0.5$\,ML; 
(c) top-{}layer-{}shift reconstruction for a H-{}coverage
$\Theta>0.5$\,ML. 
The white (shaded) circles represent the W atoms in the 
surface (subsurface) layer. 
The hydrogen positions are indicated by full dots. 
(d) Illustration of the structure parameters used in 
Table\,\protect{\ref{TStruc}}. 
The in-{}plane lattice constant is denoted by $a_0$.}
\label{FStruc}
\end{figure}
An additional puzzle to the already confusing picture is added by the 
observation of a symmetry loss in the low-{}energy electron diffraction 
(LEED) pattern of W\,(110) upon H-{}adsorption.
The phenomena was believed being caused by a H-{}induced 
displacement of the top 
layer W atoms along the $[\overline{1}10]$ direction~\cite{chun86} 
(see Figs.\,\ref{FStruc}b and c). 
For H/Mo\,(110) similar studies do not provide any evidence for
a top-{}layer-{}shift reconstruction~\cite{altm87}.

The goal of our work is to give a comprehensive explanation for the 
observed anomalous behavior and clarify the confusing picture 
drawn by the different experimental findings. 
We show that the anomalous behavior observed in the He and electron 
loss experiments is indeed governed by H-{}induced 
nesting features in the Fermi surface. 

The remainder of the paper is organized as follows. 
First, Section~\ref{SMeth} gives a brief introduction into 
the {\em ab initio} method used. 
The results of our work are presented in Section~\ref{SCalc}:
In Subsection~\ref{SAtom} we study the atomic geometry of the clean and 
H-{}covered Mo(110) and W(110) surfaces. 
In particular, we check whether the hydrogen adsorption induces a 
top-{}layer-{}shift reconstruction. 
Following in Subsection~\ref{SElec}, we focus on the electronic structure 
of those surface systems. 
In this context, we are particularily interested in how the Fermi 
surface changes upon H-{}adsorption: 
Is there evidence for Fermi-{}surface nesting? 
If yes, how does it connect to the ARP measurements of the Fermi surfaces 
and the critical wavevectors detected by HAS and HREELS?
Along the same line within the model of a Kohn anomaly, 
Subsection~\ref{SPhon} digs into the coupling between electronic 
surface states and surface phonons. 
There, we discuss the mechanism which causes the small anomaly $\omega_1$. 
In Subsection~\ref{SSusc} we examine the He scattering mechanism, 
study the spectrum of electron-{}hole excitations, and 
explain the physical nature of the huge anomalous branch $\omega_2$. 
Finally, the major aspects of our work are summarized in 
Section~\ref{SSumm}. 
The Appendix presents several detailed test calculations giving 
an insight into the properties and the accuracy of our method.  

\section{Method}\label{SMeth}
All our calculations employ the density-{}functional theory (DFT). 
They are performed within the framework of the 
local-{}density approximation (LDA) for the
exchange-correlation energy functional~\cite{cepe80,perd81}.
For the self-{}consistent solution of the Kohn-{}Sham (KS) equations
we employ a full-potential linearized augmented plane-wave (FP-{}LAPW)
code~\cite{blah85,blah90,blah93a} which we enhanced by the direct 
calculation of atomic forces~\cite{yu91,kohl96a}. 
Combined with damped Newton dynamics~\cite{stum9X} this enables an 
efficient determination of fully relaxed atomic geometries. 
Also, it allows the fast evaluation 
of phonon frequencies within a frozen-{}phonon approach. 

\subsection{LAPW Formalism}
There exists a variety of different LAPW 
programs~\cite{blah90,koel72,koel75,ande73,ande75,wimm81,jans84,matt86,sing94,blue88,sole89}. 
We therefore briefly specify the main features of our code. 
This is also in order to clarify the meaning of the LAPW parameters which 
determine the numerical accuracy of the calculations. 

In the augmented plane-{}wave method space is divided into the 
inter\-sti\-tial region (IR) and non-{}over\-lap\-ping 
muffin-{}tin (MT) spheres cen\-tered at the atomic 
sites~\cite{slat37,slat64,louc67}. 
This division accounts for the atomic-{}like character of the 
wavefunctions, the potential, and the electron density close to the nuclei 
and that the behavior of these quantities is smoother in between the 
atoms. 
The basis functions $\phi_{\bf K}({\bf r})$ which we employ in our code 
for the expansion of the electron wavefunctions of the KS equation 
\begin{equation}
\psi_{{\bf k},i}({\bf r}) = 
\sum_{|{\bf K}|\leq K^{\rm wf}} C_{i}({\bf K}) \phi_{{\bf K}}({\bf r})
\end{equation}
are defined as 
\mediumtext
\begin{equation}\label{Ewfbas}
\phi_{\bf K}({\bf r}) = 
\left\{ \begin{array}{ll}
\Omega^{-1/2} \exp(i{\bf K}{\bf r}),& {\bf r} \in {\rm IR} \\
\ \\
{\displaystyle 
\sum_{\stackrel{\scriptstyle lm}{l \leq l^{\rm wf}}} [ 
a_{lm}^I({\bf K})u_l^I(\epsilon_l^I,r_I) +
b_{lm}^I({\bf K})\dot{u}_l^I(\epsilon_l^I,r_I)
]Y_{lm}(\hat{r}_I)},& r_I \le s_I \quad.
\end{array}\right. 
\end{equation}
\narrowtext\noindent
Here, $\bf K=\bf k+\bf G$ denotes the sum of a reciprocal lattice vector 
$\bf G$ and a vector $\bf k$ within the first Brillouin zone. 
The wavefunction cutoff $K^{\rm wf}$ limits the number of 
those $\bf K$ vectors and thus the size of the basis set. 
The other symbols in eq.\,(\ref{Ewfbas}) have the following meaning: 
$\Omega$ is the unit cell volume, $s_I$ specifies the MT radius, and 
${\bf r}_I = {\bf r}-{\bf R}_I$ 
represents a vector within the MT sphere of the $I$-th atom. 
Note that $Y_{lm}(\hat{r})$ is a complex spherical 
harmonic with $Y_{l-m}(\hat{r})=(-1)^mY^*_{lm}(\hat{r})$. 
The radial functions $u_l(\epsilon_l,r)$ and 
$\dot{u}_l(\epsilon_l,r)$ are solutions of the equations
\begin{eqnarray}\label{urad}
H^{\rm sph}\,u_l(\epsilon_l,r)Y_{lm}(\hat{r}) &=& 
\epsilon_l\, u_l(\epsilon_l,r)   Y_{lm}(\hat{r})\\
\label{udot}
H^{\rm sph}\,\dot{u}_l(\epsilon_l,r) Y_{lm}(\hat{r}) &=&
 [\epsilon_l \dot{u}_l(\epsilon_l,r) + 
u_l(\epsilon_l,r)] Y_{lm}(\hat{r})
\end{eqnarray}
which are regular at the origin. 
Here, the KS operator $H^{\rm sph}$ contains only the spherical 
average of the effective potential within the respective MT. 
The expansion energies $\epsilon_l$ are chosen somewhere within the 
respective energy bands with $l$-character, whereas 
the coefficients $a^I_{lm}({\bf K})$ and $b^I_{lm}({\bf K})$ are fixed 
by requiring that value and slope of the basis functions are 
continuous at the surface of the $I$-th MT sphere.  
Obviously, this can only be fulfilled for $l \leq l^{\rm wf}$, but 
as $l^{\rm wf}$ is typically high (e.g. $l^{\rm wf} \geq 8$) this does 
not represent a problem.

The representation of the potentials and densities resembles 
the one employed for the wavefunctions, i.e.,
\begin{equation}\label{vbas}
V ({\bf r})= 
\left\{\begin{array}{ll} 
\displaystyle
\sum_{|{\bf G}|\leq G^{\rm pot}} V_{\bf G} \exp(i{\bf G}{\bf r}), 
& {\bf r} \in {\rm IR} \\ \ \\
\displaystyle\hspace*{3.5mm}
\sum_{\stackrel{\scriptstyle lm}{l \leq l^{\rm pot}}}
V_{lm,I}(r_I) Y_{lm}(\hat{r}_I),& 
r_I \le s_I\quad.
\end{array}\right.
\end{equation}
Thus, no shape approximation is introduced. The quality of this 
full-potential description is controlled by the cutoff parameter 
$G^{\rm pot}$ for the reciprocal lattice vectors $\bf G$ and 
the maximum angular momentum $l^{\rm pot}$ of the 
$(l,m)$-representation inside the MTs.

\subsection{Slab Systems and Parametrization}
The substrate surfaces are modeled by five-, seven-, and nine-{}layer slabs
repeated periodically and separated by a vacuum region whose 
thickness is equivalent to four substrate layers. 
The MT radii for the W and Mo atoms are chosen to 
be $s_{\rm Mo}=s_{\rm W}=1.27\,$\AA. 
Note that the experimental interatomic distances of bulk Mo and W are 
2.73\,\AA and 2.74\,\AA, respectively. 
For hydrogen the MT radius is set to $s_{\rm H}=0.48\,$\AA. 

In the case of the W surfaces the valence and semi-{}core 
electrons are treated scalar-{}relativistically while 
the core electrons are handled fully relativistically.  
For Mo all calculations are done non-{}relativistically.
In both metals we introduce a second energy window for the treatment 
of the semi-{}core electrons (4$s$ and 4$p$ for Mo, 5$s$ and 5$p$ for W). 
The in-{}plane lattice constants (compare Fig.~\ref{FStruc})
$a_{\rm W}^{\rm theo}=3.14\,$\AA\ and
$a_{\rm Mo}^{\rm theo}=3.13\,$\AA\ are calculated to be without 
including zero-{}point vibrations. 
They are in good agreement with the respective measured bulk lattice
parameters at room temperature and other theoretical results 
(see Appendix~\ref{ABulk}).

For the potential the $(l,m)$ representation within each MT sphere 
is taken up to $l^{\rm pot}= 4$ while the kinetic-{}energy
cutoff for the interstitial region is set to $\frac{\hbar^2}{2m}(G^{\rm wf})^2=100$\,Ry. 
Generally, we choose the plane-{}wave cutoff for the wavefunctions to be 
$\frac{\hbar^2}{2m}(K^{\rm wf})^2=12\,$Ry;
only for the frozen-{}phonon calculations we use a slightly smaller 
cutoff of $\frac{\hbar^2}{2m}(K^{\rm wf})^2=10\,$Ry which is indeed 
sufficient.  
The $(l,m)$ representation of the wavefunctions within the MTs is taken 
up to $l^{\rm wf}=8$. 
Also, we employ Fermi smearing with a broadening of 
$k_{\rm B}T_{\rm el}=68$\,meV in order to stabilize the self-{}consistency 
and the ${\bf k_{\|}}$-{}summation. 
All energies given below are however extrapolations to zero temperature 
$T_{\rm el}$~\cite{neug92,alip9X}. 
We performed systematic tests comparing the LDA and the 
generalized gradient approximation~\cite{perd92a,perd92b} 
exchange-{}correlation functionals. 
For the quantities reported in this paper (total energy differences of  
different adsorption sites, bond lengths, etc.) 
both treatments give practically the same results. 
All $\bf k$ sets used were determined by using the special $\bf k$ point 
scheme of Monkhorst and Pack~\cite{monk76}. 
Also, we checked the ${\bf k_{\|}}$-{}point convergence: 
In the case of atomic and electronic structure calculations a 
two-{}dimensional uniform  
mesh of 64 ${\bf k_{\|}}$-{}points within the $(1\times 1)$ SBZ gives stable 
results. 
For the evaluation of frozen-{}phonon energies a set 
of 56\,${\bf k_{\|}}$-{}points is employed within the SBZ of the enlarged 
$(1\times 2)$ and $(2\times 1)$ surface cells 
(see subsection~\ref{SAppPhon}). 
In Table~\ref{TParam} we summarize the main features of our LAPW parameter 
setting. 
For more details the reader is referred to the discussion in the Appendix. 
\begin{table}
\caption{LAPW parameter setting used in Section~\protect{\ref{SCalc}}
for the bulk studies, 
the calculation of the atomic and electronic surface strucutures, 
and the evaluation of frozen-{}phonons.}
\begin{tabular}{lccc}
LAPW                    &bulk   &\multicolumn{2}{c}{surface}\\
\cline{3-4}
parameter               &       &structure      &phonons\\
\hline
$l^{\rm pot}$           &4      &4              &3\\
$\frac{\hbar^2}{2m}\frac{\hbar^2}{2m}(G^{\rm pot})^2$\ [Ry]
                        &64     &100            &100\\
$l^{\rm wf}$            &8      &8              &8\\
$\frac{\hbar^2}{2m}(K^{\rm wf})^2$\ [Ry]
                        &12     &12             &10\\
Fermi smearing [meV]    &68     &68             &68\\
\hline
number of $\bf k$-points\\
~for valence electrons  &3375   &64             &56\\
~for semicore electrons &216    &9              &4\\
~in unit cell           &bulk bcc&($1\times1$)  &($1\times2$)\\
                        &       &               &($2\times1$)\\
\end{tabular}
\label{TParam}
\end{table}

\section{Results}\label{SCalc}
\subsection{Atomic Structure}\label{SAtom}
Using the directly calculated forces in combination with a damped Newton 
dynamics of the nuclei the atomic structure 
of stable and metastable geometries is obtained quite automatically. 
It is important that the starting configuration for an optimization run 
is of low (or no) symmetry. 
Otherwise the system may not relax into a low-{}symmetry structure of 
possibly lower energy. 
Also, independent of the optimization method used, there is 
always a chance that the minimization of the total energy 
leads into a local minimum instead of {\em the} global one. 
In order to reduce and hopefully abolish this risk one has to conduct 
several optimizations with strongly varying starting configurations. 
We consider a system to be in a stable (or meta-{}stable) 
geometry if all force components are smaller than 
20\,meV/\AA.
The error in the structure parameters of the relaxed system 
is then $\pm 0.02\,\mbox{\AA}\approx\pm 1.0\%d_0$ for the W systems and 
$\pm 0.01\,\mbox{\AA}\approx\pm 0.5\%d_0$ for the Mo surfaces 
(see Appendix~\ref{ASurf}). 

\widetext 
\begin{table}
\caption{Calculated relaxation parameters for the clean and H-covered
(110) surfaces of Mo and W. The height of the hydrogen above the
surface and its $[\overline{1}10]$ offset from the [001] bridge position
are denoted by $d_{\rm H}$ and
$y_{\rm H}$, respectively. 
The shift of the surface layer with respect to the substrate
is $y_1$.
The parameters $\Delta d_{ij}$ describe the percentage change of the
interlayer distance between the $i$-th and the $j$-th substrate layers with
respect to the bulk interlayer spacing $d_0$.
For each system the results of a five- , seven- , and nine-{}layer 
calculation as well as of a recent LEED 
analysis~\protect{\cite{arno96a,arno96b}} are presented 
(labelled as ``5'', ``7'', ``9'', and ``L'').}
{\small
\begin{tabular}{lccccccc}
system
&&$y_{\rm H}$
&$d_{\rm H}$
&$y_1$
&$\Delta d_{12}$
&$\Delta d_{23}$
&$\Delta d_{34}$        \\
&&$[$\AA$]$
&$[$\AA$]$
&$[$\AA$]$
&$[\%d_0]$
&$[\%d_0]$
&$[\%d_0]$              \\
\hline
Mo(110)
&5
&$-$
&$-$
&$-$
&$-$5.9
&$-$0.8
&$-$
\\
&7
&$-$
&$-$
&$-$
&$-$4.5
&+0.5
&0.0
\\
&9
&$-$
&$-$
&$-$
&$-5.0$
&$+0.7$
&$-0.3$
\\
&L
&$-$
&$-$
&$-$
&$-4.0\pm0.6$
&$+0.2\pm0.8$
&$0.0 \pm1.1$
\\
\hline
H/Mo(110)
&5
&0.63   &1.08   &0.05   &$-$2.7 &$-$0.4 &$-$
\\
&7
&0.60   &1.07   &0.03   &$-$2.1 &$+$0.1 &$-$0.1
\\
&9
&$0.62$
&$1.09$
&$0.04$
&$-2.5$ %3.3!!!
&$+0.3$
&$-0.2$
\\
&L
&$0.55\pm 0.4$
&$1.3\pm 0.3$
&$0.0\pm 0.1$
&$-2.0\pm 0.4$
&$0.0\pm0.5$
&$0.0\pm0.8$
\\
\hline
W(110)
&5          &$-$    &$-$    &$-$    &$-$4.1 &$-$0.2 &$-$    \\
&7          &$-$    &$-$    &$-$    &$-$3.3 &$-$0.1 &$-$0.4 \\
&9
&$-$
&$-$
&$-$
&$-$3.6
&$+$0.2
&$-$0.3
\\
&L
&$-$
&$-$
&$-$
&$-3.1\pm 0.6$
&$0.0\pm0.9$
&$0.0\pm1.0$
\\
\hline
H/W(110)
&5        &0.68   &1.12   &$0.01$ &$-$1.4 &0.0    &$-$    \\
&7                &0.67   &1.11   &$0.05$ &$-$1.3 &$+$0.3 &+0.3   \\
&9
&0.67
&1.09
&0.02
&$-1.4$
&$-0.3$
&$-0.1$
\\
&L
&$0.56\pm0.4$
&$1.2\pm0.25$
&$0.0\pm0.1$
&$-1.7\pm0.5$   
&$0.0\pm0.6$
&0.0$\pm$0.9
\end{tabular}}
\label{TStruc}
\end{table}
\narrowtext
The relaxation parameters calculated for the clean and H-{}covered 
(110) surfaces of Mo and W are presented in Table~\ref{TStruc}
(Fig.~\ref{FStruc}d serves to clarify the meaning of these parameters). 
The converged nine-{}layer-{}slab results are remarkably similar for 
both transition metals. 
Moreover, they show excellent quantitative agreement with the results 
of a recent LEED analysis also presented in Table~\ref{TStruc}.
For a detailed comparision between experiment and theory we 
refer to Ref.~\onlinecite{arno96a}. 
As the energetically most favorable hydrogen adsorption site on both 
W\,(110) and Mo\,(110) we identify a quasi-{}threefold position 
(indicated as ``H'' in Fig.~\ref{FStruc}a). 
The calculated adsorption energies for other possible positions 
shown in Fig.~\ref{FStruc}a 
are several 100\,meV less favorable~\cite{kohl95a}. 

\begin{figure}
\begin{center}
\hspace{.1mm}
\psfig{figure=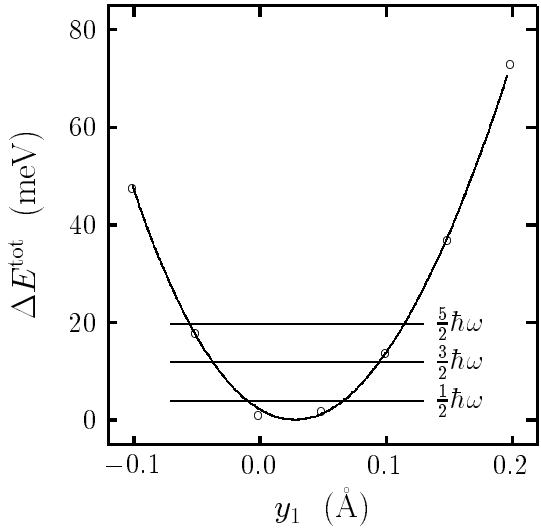,width=7cm}
\vspace{2mm}
\end{center}
\caption{Change of total energy $E^{\rm tot}$ versus 
top-{}layer-{}shift $y_1$ calculated for a H/\,W(110) seven-{}layer 
slab system. 
For each data point the whole surface is relaxed keeping only 
the substrate $[1\overline10]$-{}coordinates $y_1$ and $y_2=y_3=0\,$\AA~ 
fixed.
Also shown are the first three oscillator eigenstates calculated 
from a harmonic expansion of the total energy $E^{\rm tot}(y_1)$.}
\label{FShift}
\end{figure}
Our investigations also throw light on the suggested 
model of a H-{}induced structural change: 
For both materials the calculated shift $y_1$ is only of the order of 
0.01\,\AA\ and thus there is no evidence for a pronounced 
top-{}layer-{}shift reconstruction. 
Moreover, this subtle change in the surface geometry is unlikely to 
be resolved experimentally due to zero-{}point vibrations. 
This aspect becomes evident by evaluating the total energy 
with respect to a rigid top-{}layer-{}shift. 
In Fig.~\ref{FShift} we present the results of such a calculation 
for H/W\,(110) and depict the first three vibrational eigenstates 
obtained from a harmonic expansion of the total energy $E^{\rm tot}(y_1)$; 
for H/Mo\,(110) the energetics is similar. 
Since we have $k_{\rm B}T\approx 25$\,meV at room temperature 
thermal fluctuations of $y_1$ are of the order of 0.1\,\AA. 
This is considerably larger than the theoretically predicted 
ground state value of $y_1$. 
In Ref.~\onlinecite{arno96a} Arnold {\em et al.}  offer an explanation 
of how the previous LEED work was probably mislead by assuming that 
scattering from the H layer is negligible. 

\subsection{Electronic Structure}\label{SElec}
In Section~\ref{SIntro} we already pointed out that one needs to 
find pronounced nesting features in the Fermi surface of the 
H-{}covered surfaces in order to make the model of a Kohn anomaly work. 
After the determination of the relaxed surface configurations we 
therefore turn our attention towards the electronic structure 
of those systems. 
In particular, we focus on surface states close 
to the Fermi level. 

One should keep in mind that DFT is not expected to give 
the exact Fermi surface, i.e., the self-{}energy operator may 
have (a strong or weak) $\bf k$ depencence, different from that 
of the exchange-{}correlation potential. 
Moreover, we encounter additional problems because it is rather 
difficult in slab systems to distinguish between surface resonances 
and pure surface states. 
Due to interactions between the two faces on either side of the 
slab resonances can be shifted into a bulk band gap. 
There, they are easily mixed up with real surface states. 
By contrast, in  a semi-{}infinite substrate 
only true surface states are localized in the bulk band gap and 
decay exponentially into the bulk. 
Therefore, our slab method lacks a clear indicator for the 
characterization of electronic states. 
However, we found that the localization $w^{\rm MT}$ 
within the MT spheres of the top two surface layers 
of a seven-{}layer slab is a useful and suitable measure 
($60\%< w^{\rm MT}$: surface-{}state like, 
$30\%< w^{\rm MT} < 60\%$: surface-{}resonance like,
$w^{\rm MT}< 30\%$: bulk like).
Nevertheless, this approach is anything but exact. 
Also, the calculated $\bf k_{\|}$ space location is only accurate 
if the respective band is strongly localized at the surface.

Experimentalists face similarly serious problems: 
In ARP the Fermi surface is determined by extrapolation peaks found 
below the Fermi level. 
However, the value of the Fermi energy itself 
is only known within an uncertainty of about $\pm100\,$meV. 
This uncertainty can amount to a significant error in the extrapolated 
$\bf k_{\|}$ space position of states, especially if the 
respective bands are flat~\cite{chun92}. 

\begin{figure}
\begin{center}
\hspace{.1mm}
\psfig{file=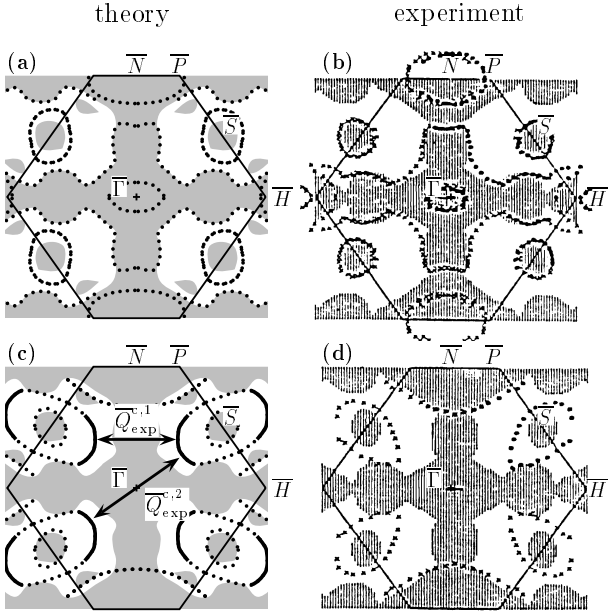,width=8cm}
\vspace{2mm}
\end{center}
\caption{Theoretical and ARP Fermi surfaces of the clean 
(upper part) und the H-{}covered (lower part) W(110) surface. 
The solid (dotted) lines denote surface resonances or surface 
states which are localized by more than 60 \% (30 \%) in the MTs 
of the two top W layers. 
Shaded areas represent the (110) projected theoretical W bulk 
Fermi surface. 
The arrows $\overline Q^{\rm c,1}_{\rm exp}$ and 
$\overline Q^{\rm c,2}$ are the critical wave vectors found in 
HAS. 
The ARP data stem from Refs.~\protect{\onlinecite{gayl89,jeon89b}}. }
\label{FFermiW}
\end{figure}
\begin{figure}
\begin{center}
\hspace{.1mm}
\psfig{file=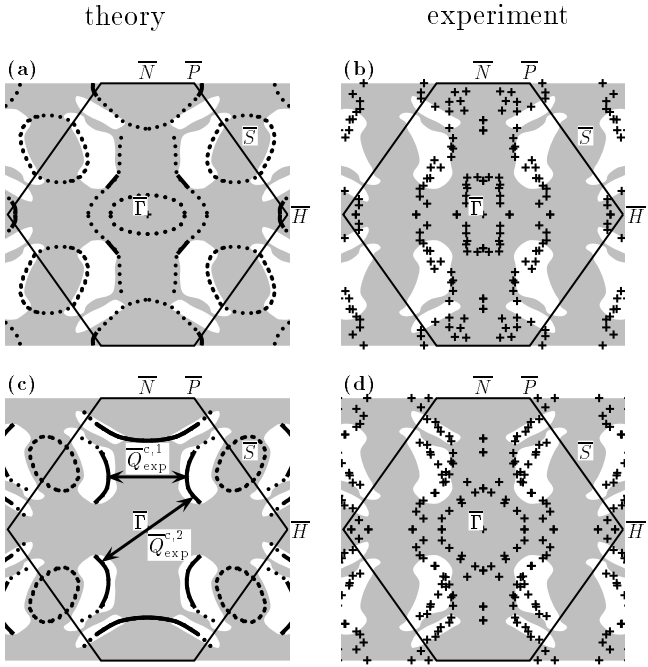,width=8.5cm}
\vspace{2mm}
\end{center}
\caption{Theoretical Fermi surfaces of the (a) clean 
and (c) H-{}covered Mo\,(110) surface. 
Also presented are data points (+) 
which stem from an ARP study of the (b) clean and (d) H-{}covered 
Mo$_{0.95}$Re$_{0.05}$\,(110) surface~\protect{\cite{okad96}}.
The presentation is equivalent to Fig.~\protect{\ref{FFermiW}}.}
\label{FFermiMo}
\end{figure}
In Figs.~\ref{FFermiW} and \ref{FFermiMo} we report the calculated 
Fermi surfaces and compare them to experimental ones obtained by 
Kevan's~\cite{gayl89,jeon89b} and Plummer's~\cite{okad96} groups. 
Let us first focus on the theoretical results which are --- as 
in the case of the surface geometries --- again very similar 
for Mo and W. 

\begin{figure}
\begin{center}
\hspace{.1mm}
\psfig{figure=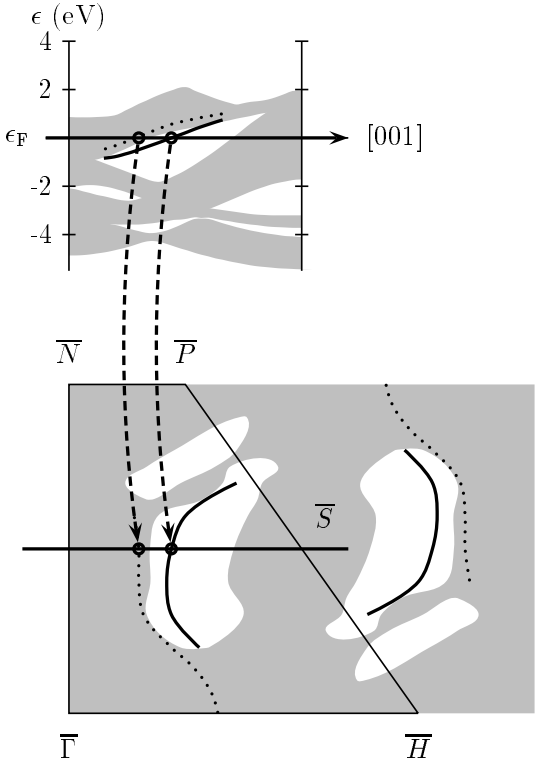,width=7cm}
\vspace{2mm}
\end{center}
\caption{Hydrogen induced shift of the ($d_{3z^2-r^2},d_{xz}$) band 
at the Mo(110) surface. 
The upper part of the figure represents a one-{}dimensional cut through 
the two-{}dimensional surface bandstructure. 
Shaded areas illustrate the (110) projected bulk band structure. 
The dotted (solid) lines depict the position of the ($d_{3z^2-r^2},d_{xz}$) 
band for the 
clean (H-{}covered) surface. 
The arrows indicate how the bandstructure
translates into the two-{}dimensional Fermi surface shown in the lower 
part. 
In Fig.~\protect{\ref{FDensPlot}} the charge density of the 
two states marked by the arrows is presented.} 
\label{FBand}
\end{figure}
For both systems the H-{}adsorption 
induces the shift of a band with $(d_{3z^2-r^2},d_{xz})$ 
character to lower binding energies~\cite{kohl95a,rugg95}. 
This effect which is illustrated in Fig.~\ref{FBand} moves the 
Fermi line associated with this band into the band 
gap of the surface projected band structure. 
Subsequently, the respective states become true surface 
states. 
It is important to note that the band shift is due not 
to a hybridization between hydrogen orbitals 
and ($d_{3z^2-r^2},d_{xz}$) bands but to a hydrogen induced 
modification of the surface potential. 
The bonding states of the hydrogen-{}substrate interaction are about 
5\,eV below the Fermi energy and the anti-{}bonding 
states are 4\,eV above. 
Therefore, they are not involved in this process which takes place 
at the Fermi level. 

\begin{figure}
\begin{center}
\hspace{.1mm}
\psfig{figure=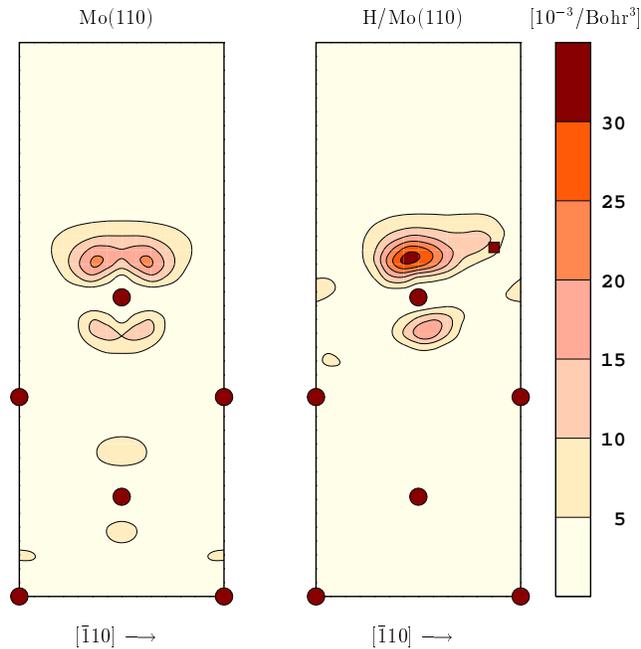,width=9cm}
\vspace{2mm}
\end{center}
\caption{
Charge density plot along the $(010)$ plane for Mo(110) and 
H/Mo(110): 
Shown are the two states marked in Fig.~\protect{\ref{FBand}} by 
arrows. 
Only the plane-{}wave part of the wavefunctions is shown. 
The positions of the Mo atoms are marked by dots. 
The square represents the hydrogen atom in the hollow adsorption 
site.}
\label{FDensPlot}
\end{figure}
\begin{table}
\caption{Theoretical Fermi surface nesting vectors compared to critical 
wavevectors obtained by HAS and HREELS experiments
\protect{\cite{hulp92a,lued94,bald94a,bald94b}}.}
\begin{tabular}{llcc}
direction       &system         &\multicolumn{2}{c}{
                                $|{\overline Q}^{\rm c}|\ \ 
                                [\mbox{\AA}^{-1}]$}\\
                                \cline{3-4}
                &               &theory &experiment\\
\hline  $\overline{\Gamma H}\ \ $
                        &H/Mo(110)      &0.86   &0.90 \\
                        &H/W(110)       &0.96   &0.95 \\
\hline  $\overline{\Gamma S}\ \ $
                        &H/Mo(110)      &1.23   &1.22\\
                        &H/W(110)       &1.22   &1.22\\
\end{tabular}
\label{TNest}
\end{table}
The shifted $(d_{3z^2-r^2},d_{xz})$ band is characterized by a high 
density of states at the Fermi level. 
For the clean Mo(110) surface one finds that the 
($d_{3z^2-r^2},d_{xz}$) 
band has a MT localization of $w^{\rm MT}\approx 30\%$. 
Due to the H-{}adsorption this value increases to more than 60\%. 
This modification is also visible in the charge density
plots in Fig.~\ref{FDensPlot}. 
More important, the new Fermi contour reveals pronounced nesting 
features. 
In Figs.~\ref{FFermiW}c and \ref{FFermiMo}c 
one sees the dramatic changes due to the H-induced shift. 
Segments of the two-{}dimensional Fermi contour of the 
($d_{3z^2-r^2},d_{xz}$) band are now running parallel to 
$\overline{\Gamma N}$ and normal to $\overline{\Gamma S}$. 
There are two nesting vectors 
$\overline Q^{\rm c,1}$ and $\overline Q^{\rm c,2}$ 
which connect those segments in different parts of the SBZ. 
As can be seen in Table~\ref{TNest} they agree very well with the 
HAS measured critical wavevectors. 

With respect to those nesting vectors our results contradict 
the photoemission studies by Kevan and 
coworkers~\cite{gayl89,jeon89a,jeon89b,jeon88} which were 
depicted in Figs.~\ref{FFermiW}b,d (for W) 
and Fig.~2 in Ref.~\onlinecite{kohl95a} (for Mo). 
Apart from that Kevan's and our data compare rather well. 
Therefore, it is difficult to give a plausible reason 
for the discrepancies. 
However, in view of the fact that for the clean surface the 
theoretical and experimental Fermi surfaces are in very good 
agreement and that our calculated Fermi nesting vectors agree 
very well with the HAS and HREELS anomalies we 
dared to suggest that those differences may be due to 
the already discussed problems within the experimental 
analysis~\cite{kohl95a}. 
We also note that recent ARP studies~\cite{okad96} of a related 
system which we present in Fig.~\ref{FFermiMo} seem to support our 
conclusion. 
This experimental work deals with the (110) surface 
of the alloy Mo$_{0.95}$Re$_{0.05}$ but the surface physics of 
Mo$_{0.95}$Re$_{0.05}$\,(110) and Mo\,(110) 
should be practically the same because in both cases the top 
layer consists only of Mo atoms. 
This assumption is also backed by test calculations where we 
simulated a Mo$_{0.95}$Tc$_{0.05}$ alloy be the virtual 
crystal approximation and found only 
minor quantitative changes. 

From Figs.~\ref{FFermiMo}c and \ref{FFermiMo}d it becomes clear 
that in particular for the important $(d_{3z^2-r^2},d_{xz})$ surface 
band, which was not seen by Kevan's group, experiment and DFT now 
agree very well. 
There are however still differences: 
Theory predicts bands centered at $\overline S$ which are not seen by 
ARP whereas the band circle at $\overline \Gamma$ in Fig.~\ref{FFermiMo}d 
is only observed experimentally. 
Also, in the calculations we find an elliptical band centered at the 
$\overline \Gamma$ point whereas in ARP the same band has the shape 
of a rectangle. 
Those discrepancies are probably due to matrix elements of the photoemission 
process and to the inherent inaccuracies in theory and experiment which 
we mentioned above.  
Nevertheless, the theoretical results are encouraging because they 
provide strong evidence for a link between the HAS anomalies 
and the Fermi surfaces.

\subsection{Vibrational Properties}\label{SPhon}
The study of the electronic structure revealed pronounced nesting 
features for the H-{}covered surfaces. 
At this point, the following question arises: 
How does the surface react to this apparent electronic
instability? 
At the $\bf k$ vectors of the Fermi surface nesting the coupling between 
electrons and phonons is expected to become significant which implies 
a possible breakdown of the Born-{}Oppenheimer approximation. 
This leads to a softening of the related phonons. 
If the electron-{}phonon coupling is strong and the energetic cost 
of a surface distortion is small the nesting could even trigger 
a reconstruction combined with the build-{}up of a charge-{}density 
wave as in the case of the (001) surfaces of W~\cite{erns92a} and 
Mo~\cite{hulp89}. 
Then, at the reconstructed surface, the Born-{}Oppenheimer approximation 
is valid again. 
It is clear that one needs to perform frozen-{}phonon calculations in 
order to determine the actual strength of the electron-{}phonon 
coupling and the resulting phonon softening~\cite{kohl96b}.  

The experimental vibrational spectra of the (110) surfaces of Mo and W 
show two distinct reactions to the adsorption of hydrogen. 
Along $\overline{\Gamma H}$ and $\overline{\Gamma S}$ the H-{}adsorption
induces a softening of the Rayleigh and (to a smaller amount) of the 
longitudinal wave~\cite{bald95b} while a stiffening of these modes is 
observed along $\overline{\Gamma N}$. 
Our goal is to investigate both effects theoretically. 

We use an enlarged surface unit cell together with
a five layer slab. 
The plane-wave cutoff for the wavefunctions is reduced to 
$\frac{\hbar^2}{2m}(K^{\rm wf})^2=10$\,Ry 
(see test caculations in the Appendix~\ref{APhon}).
The geometries are defined by the five layer relaxation parameters
presented in Table \ref{TStruc}.
Some relaxation parameters, i.e., $\Delta d_{ij}$, change considerably 
when we perform a calculation with a slab of five instead of 
seven or nine metal layers. 
We note, however, that the values of the critical wavevectors 
and the position of the hydrogen with respect to the substrate surface 
are practically insensitive to the thickness of the slab. 
This indicates that the physical properties we are interested in, 
e.g. the nesting features, are well localized surface phenomena 
and that the results of our five layer slab studies can be trusted. 
As shown in Appendix~\ref{APhon} for the $\overline S$ point 
phonon the calculated frequencies are relatively insensitive to the 
the SBZ $\bf k_{\|}$ point sampling, and we conclude that 
a uniform mesh of 56 points is sufficient for the present study. 

Being zone boundary phonons 
in the Rayleigh waves at $\overline S$ and $\overline N$ 
the displacements of the surface atoms are along the direction normal 
to the surface. Moreover, the vibrations are strongly localized in the 
surface region. We assume that even with hydrogen adsorbed the coupling to modes 
parallel to the surface is small. 
Therefore, we can confine our study 
to the vibrational components normal to the surface.
Besides the first substrate layer we also include the vibrations of the 
second layer in our calculations. 
Due to the small mass the hydrogen atom follows the substrate vibrations 
adiabatically. 
In the calculations this adsorbate relaxation lowers the distortion 
energy by several meV and thus reduces the phonon energy significantly. 
Therefore, all phonon frequencies are calculated for a fully relaxed 
hydrogen adsorbate. 

\begin{figure}
\begin{center}
\hspace{.1mm}
\psfig{file=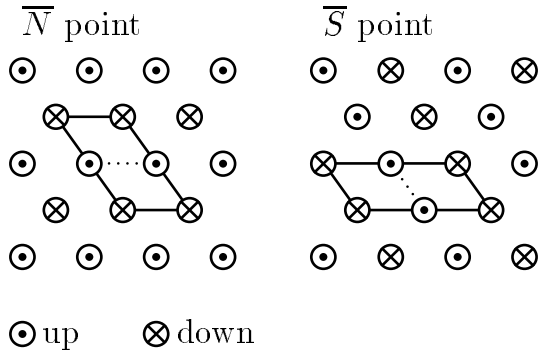,width=7cm}
\vspace{2mm}
\end{center}
\caption{Distortion pattern of the atoms of the top metal layer for 
the Rayleigh phonons at the symmetry points $\overline S$ and 
$\overline N$.}
\label{FPattern} 
\end{figure}
For the study of the vibrational properties of the systems we displace the 
substrate atoms in the surface (according to Fig.~\ref{FPattern}), relax the 
hydrogen atom, and calculate the resulting atomic forces. 
The same is done for the atoms in the subsurface layer. 
From a third-order fit to the calculated forces for five different displacement 
steps between 0\% and 5\% of the lattice constant we obtain the 
matrix elements of the dynamical matrix and then via diagonalization 
the respective phonon energies. 
The calculated frequencies are collected in Table~\ref{Tphonon}.
At the $\overline{N}$ point our results reproduce the experimentally
observed increase of the Rayleigh-{}wave frequency as hydrogen is
adsorbed.
\begin{table}
\caption{
Comparison of calculated frozen-{}phonon energies and
experimental values obtained by HAS~\protect{\cite{hulp92a,lued94}} and 
HREELS~\protect{\cite{bald94b}}.
The theoretical phonon energies are obtained using a five-{}layer slab.  
Their numerical accuracy is about $\pm 5\%$, i.e., $\pm 1$\,meV.}
\begin{tabular}{llll}
phonon          &system         &\multicolumn{2}{c}{$E^{\rm ph}$\ \ [meV]}\\
                                \cline{3-4}
                &               &theory &experiment\\
\hline
$\bar N$       &W\,(110)         &15.4   &14.5\\
                &H/W\,(110)       &17.6   &17.0\\
\hline
$\overline{S}$       &Mo\,(110)        &22.7   &$\sim$21\\
                &H/Mo\,(110)      &17.2   &$<$16\\
                \cline{2-4}
                &W\,(110)         &18.3   &16.1\\
                &H/W\,(110)       &12.0   &11.0\\
\end{tabular}
\label{Tphonon}
\end{table}

\begin{figure}
\begin{center}
\hspace{.1mm}
\psfig{file=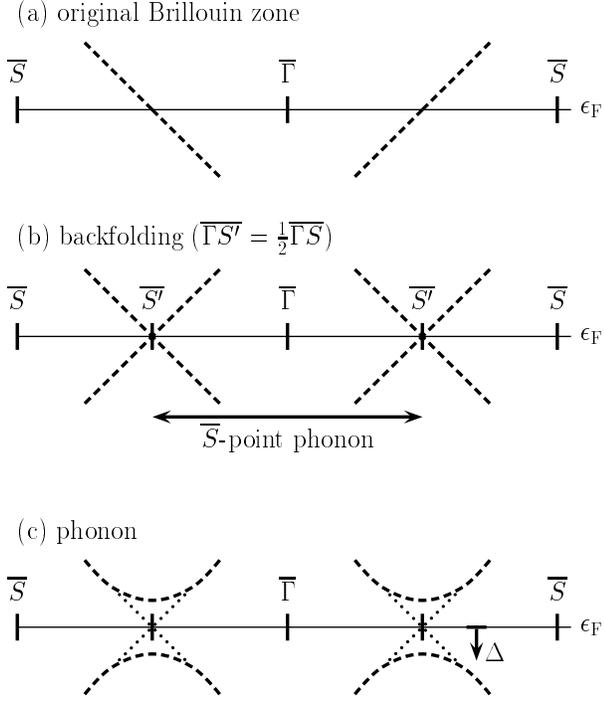,width=8cm}
\vspace{2mm}
\end{center}
\caption{Schematic representation of the mechanism 
which causes the Kohn anomaly of the H/Mo(110) and H/W(110). 
Shown is the bandstructure along \protect{$\overline{S\Gamma S}$} parallel 
to the nesting vector $\overline Q^{c,2}$ in 
Figs.~\protect{\ref{FFermiW}c} and \protect{\ref{FFermiMo}c}. 
The form of the \protect{($d_{3z^2-r^2},d_{xz}$)} band is 
indicated by dashed lines.
See text for a detailed description.}
\label{FKohn}
\end{figure}
At the ${\overline S}$ point we find that the strong coupling to 
electronic states at the Fermi level leads 
to a lowering of the phonon energy in good agreement 
with the experimental results. 
In Figs.~\ref{FKohn}a-c we schematically illustrate the mechanism 
which is responsible for this effect. 
It is, in fact, a text-{}book example of a Kohn 
anomaly due to Fermi-{}surface nesting~\cite{migd58,kohn59}: 
(a) Within the unperturbated system the ($d_{3z^2-r^2},d_{xz}$) band 
cuts the Fermi level exactly midway between $\overline \Gamma$ and 
$\overline S$. 
(b-c) 
The nuclear distortion associated with the $\overline S$-{}point phonon 
modifies the surface potential and hence removes the degeneracy at 
the backfolded zone boundary point $\overline{S'}$. 
\begin{figure}
\begin{center}
\hspace{.1mm}
\psfig{file=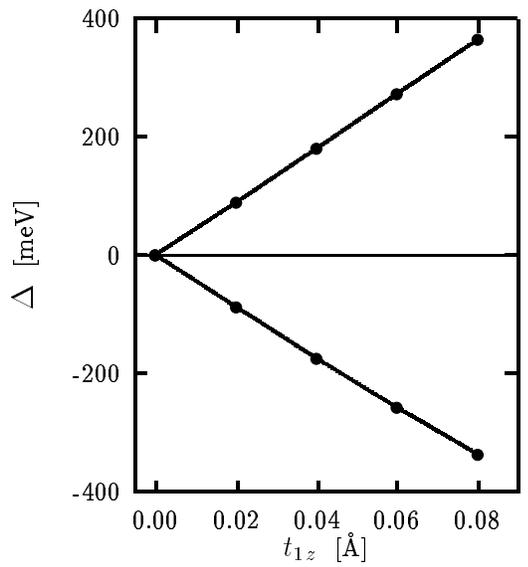,width=7cm}
\vspace{2mm}
\end{center}
\caption{H/W(110): Band gap $\Delta$ induced by a $\overline S$ point 
frozen-{}phonon distortion. 
Shown are the KS eigenvalues 
of the \protect{($d_{3z^2-r^2},d_{xz}$)} band at the $\overline{S'}$ 
point versus the amplitude of the atomic distortion 
$t_{1z}$ of the first substrate layer.}
\label{FPeierlsGap}
\end{figure}
The occupied states are shifted to lower energies 
(see Fig.~\ref{FPeierlsGap}). 
This amounts to a negative contribution of the electronic band 
structure energy to the total energy and thus a lowering of the 
phonon energy. 

A frozen-{}phonon study for the second nesting vector 
along $\overline{\Gamma H}$ is not performed because the respective 
$\overline Q^{\rm c,1}$ is highly non-{}commensurate. 
Thus, such a calculation would be very expensive.  
However, since the character of the $(d_{3z^2-r^2},d_{xz})$ band 
does not change when shifting from the $\overline{\Gamma S}$ to 
the $\overline{\Gamma H}$ nesting we expect similar results; 
this was recently confirmed by Bungaro~\cite{bung95} within the 
framework of a perturbation theory. 

The calculation of the electron-{}phonon interaction
at the Mo\,(110) and W\,(110) surfaces and its change 
due to hydrogen adsorption pinpoints the phonon character of the 
small anomaly $\omega_2$ and  identifies  the interplay between 
the electronic structure and the vibrational spectra of the 
transition metal surfaces. 
Thus, our results clearly support the interpretation that the small 
dip observed by both HAS and HREELS is due to a Kohn anomaly.
Furthermore, we find that the electron-{}phonon coupling 
is not strong enough to induce a stable reconstruction. 
Thus, the system remains in a somewhat limbo-{}like state. 

\subsection{Electron-{}Hole Excitations}\label{SSusc}
For the deep and narrow anomaly $\omega_1$ the interpretation  
appears to be less straightforward. 
It is particularly puzzling that it is only seen in the HAS 
spectra and not in HREELS. 
This difference. together with the above noted ``limbo state'' 
of the surface yield the clue to our present interpretation. 
At first, it is necessary to understand the nature of rare-{}gas 
atom scattering. 

\begin{figure}
\begin{center}
\hspace{.1mm}
\psfig{file=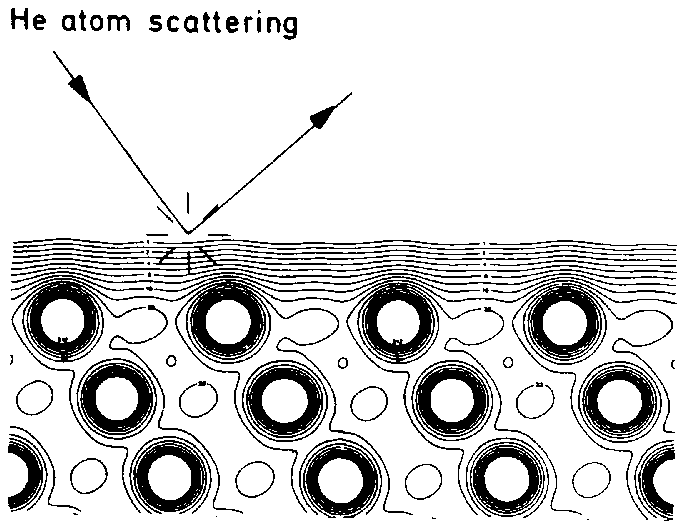,width=8cm}
\psfig{file=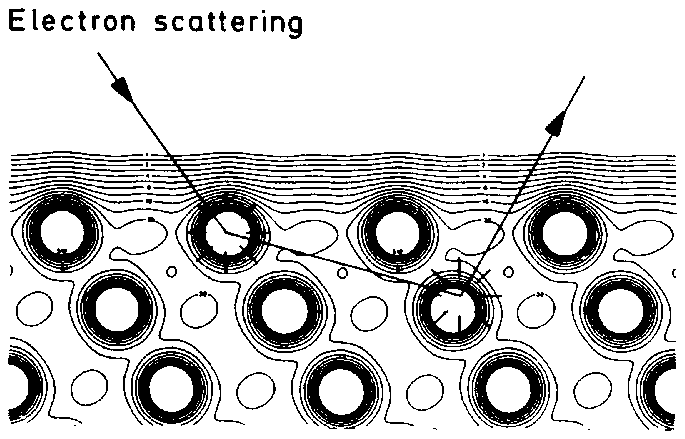,width=8cm}
\vspace{2mm}
\end{center}
\caption{Schematic illustration of the scattering 
of He atoms (HAS, upper part) and electrons 
(HREELS, lower part) at a metal surface 
(from Ref.~\protect{\onlinecite{toen91}}).}
\label{FHAS+EELS}
\end{figure}
In a recent study we found that those scattering processes 
are significantly more complicated (and more interesting)
than hitherto assumed~\cite{pete96}: 
The reflection of the He atom happens in front of the surface
at a distance of 2--3\,\AA. 
This is illustrated in Fig.~\ref{FHAS+EELS}. 
More important, it is not the {\em total} electron density of the 
substrate surface which determines the interaction but the electronic 
wavefunctions close to the Fermi level.  
In the case of the H/W\,(110) and H/Mo\,(110) systems 
it is thus plausible to assume that the He atom couples directly 
to the $(d_{3z^2-r^2},d_{xz})$ surface states mentioned above and 
excites electron-{}hole pairs. 
By contrast, the electrons in HREELS scatter at the atomic cores 
and interact only weakly with the electron density at the surface. 

In order to analyze the spectrum of those excitations as seen by 
HAS in some more detail we evaluate the local density of 
electron-{}hole excitations 
\begin{eqnarray}
\nonumber
P({\bf q_{\|}},\hbar\omega) &=&
\int_{\rm SBZ} d{\bf k_{\|}}\
w_{\bf k_{\|}+\bf q_{\|}} w_{\bf k_{\|}}\
(f_{\bf k_{\|}+\bf q_{\|}}-f_{\bf k_{\|}})\\
&&\hspace{1cm} \times \delta\left( \epsilon_{\bf k_{\|}+\bf q_{\|}}
-\epsilon_{\bf k_{\|}}-\hbar\omega\right)
\end{eqnarray}
which is a measure of the probability that a He atom looses the energy 
$\hbar \omega$ and the momentum $\bf q_{\|}$ when interacting with the 
surface elctron density. 
In our approach we use {\em all} eigenvalues $\epsilon_{\bf k_{\|}}$ and 
occupation numbers $f_{\bf k_{\|}}$ obtained via a nine-{}layer 
slab calculations. 
In order to refer to HAS we take into account the localization 
$w_{\bf k_{\|}}$ of the respective state at a distance of 2.5\,\AA\ 
in front of the substrate surface. 

\begin{figure}
\begin{center}
\hspace{.1mm}
\psfig{file=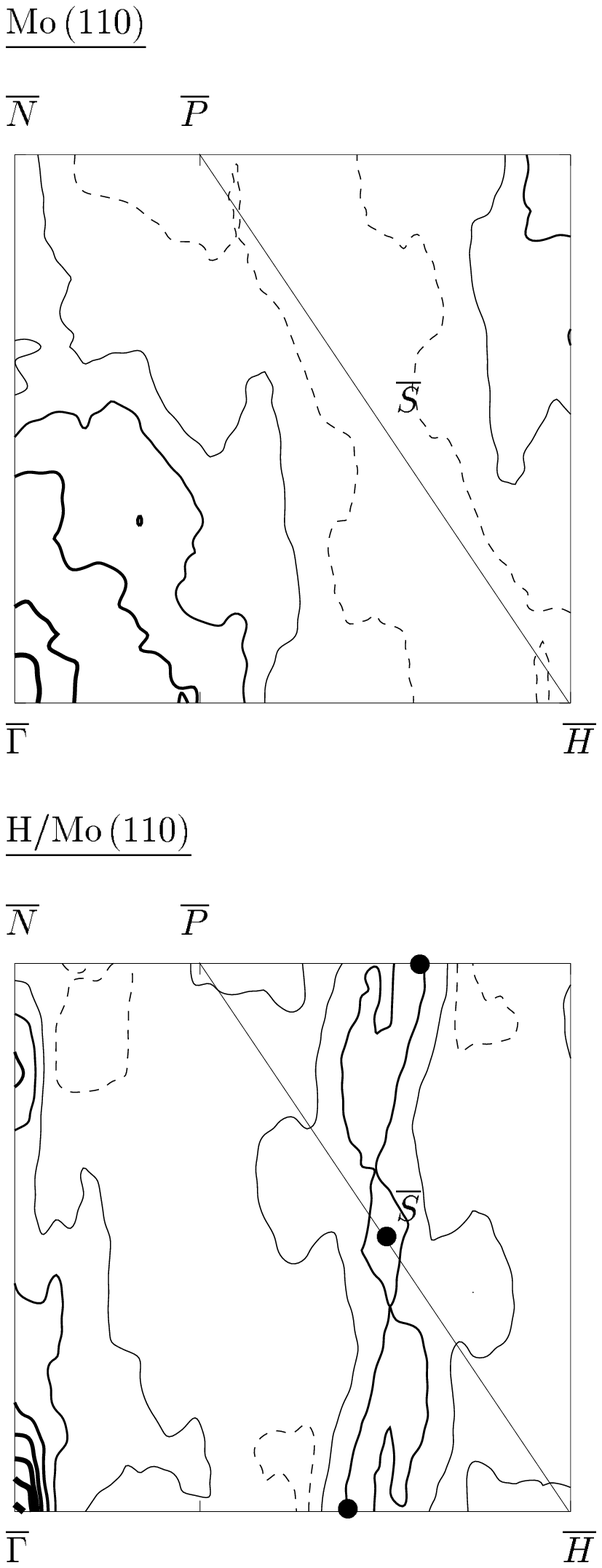,width=6cm}
\vspace{2mm}
\end{center}
\caption{Contour plot of the local probability function 
$P({\bf q_{\|}},\hbar \omega)$ of Mo\,(110) and H/Mo\,(110) 
calculated for $\hbar \omega = 0.5$\,mRy $\approx 6.8$\,meV. 
The dashed line represents a value of 3\,arbitrary units while each full 
line denotes an additional increase by 1\,arbitrary units 
(with increasing line width).
The positions of the HAS measured anomalies within 
the SBZ are indicated by dots.}
\label{FSuscMo}
\end{figure}
\begin{figure}
\begin{center}
\hspace{.1mm}
\psfig{file=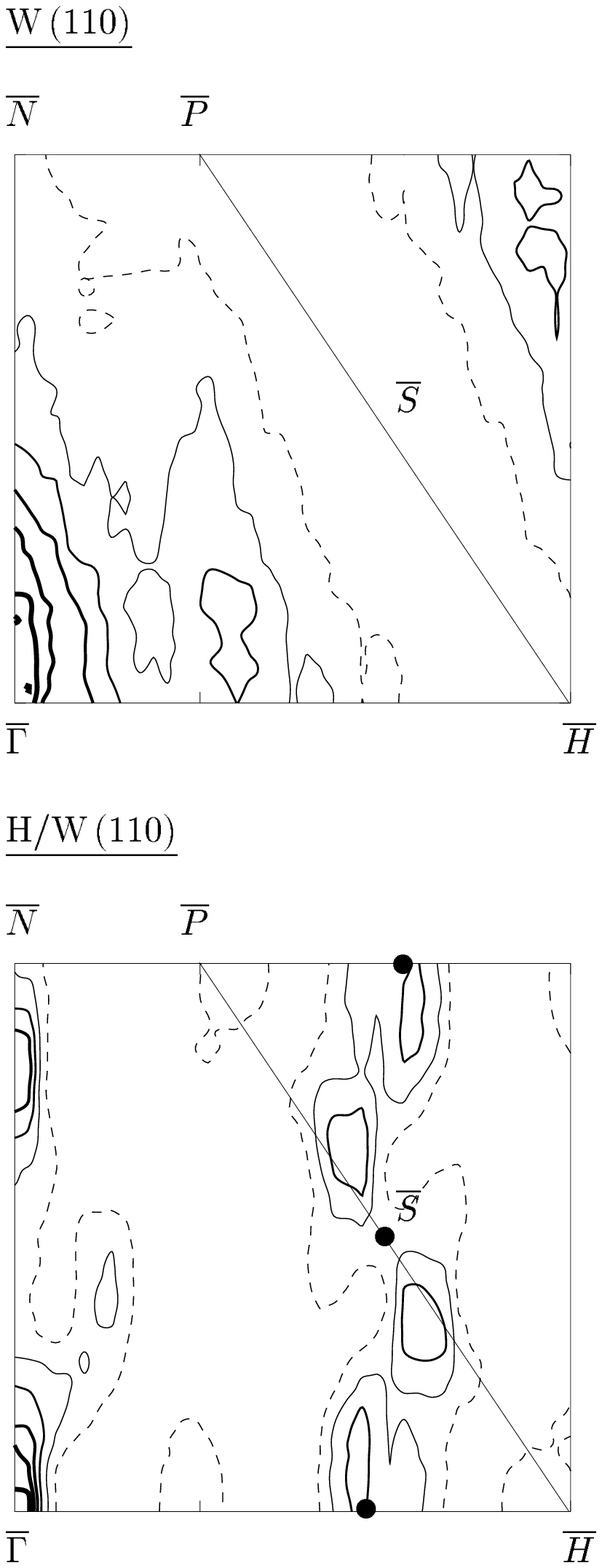,width=6cm}
\vspace{2mm}
\end{center}
\caption{Contour plot of the local probability function 
$P({\bf q_{\|}},\hbar \omega)$ of W\,(110) and H/W\,(110) 
calculated for $\hbar \omega = 0.5$\,mRy $\approx 6.8$\,meV. 
Notation equivalent to Fig.~\protect{\ref{FSuscMo}}.}
\label{FSuscW}
\end{figure}
The results obtained for Mo\,(110) and H/Mo\,(110) are presented in 
Figs.~\ref{FSuscMo}.
For the clean surface we find that the intensity of 
the electron-{}hole excitations decreases continuously 
as we move away from the $\overline \Gamma$ point towards the zone 
boundaries. 
This relatively smooth behavior of the susceptibility 
is modified considerably when hydrogen is adsorbed on the clean 
surface: 
Pronounced peaks appear, in particular one which stretches from 
the $\overline{\Gamma H}$ line through the SBZ to the symmetry point 
$\overline S$. 
This is in excellent agreement with the HAS results 
(see also Fig.~\ref{FLocus}). 
Also, there is a peak located at the upper half of the 
$\overline{\Gamma N}$ symmetry line.
It cannot be 
associated with an anomaly in the HAS spectrum. 
This result may be due to the fact that selection 
rules or matrix elements for the interaction between the 
scattering He atoms and the electron-{}hole excitations are not considered 
in $P({\bf k},\hbar \omega)$. 
In Fig.~\ref{FSuscW} the respective data for the 
clean and H-{}covered W\,(110) surfaces are displayed.  
In general, our findings are in agreement with the experimental results 
of L\"udecke and Hulpke that the anomaly stretches through the 
SBZ. 
However, we do not support the idea of a fully one-{}dimensional 
nesting mechanism with a constant [001]-{}component of the critical 
wave vector as depicted in Fig.~\ref{FLocus}.

In view of the simplicity of our approach, 
the calculated spectrum of electron-{}hole excitations 
as predicted by $P({\bf q},\hbar \omega)$
seems to be well-{}connected to the HAS measurements. 
Moreover, it is surprising how detailed the Fermi-{}surface nesting 
(compare to Figs.~\ref{FFermiW}c and \ref{FFermiMo}c) is 
still present at a (rather large) distance from the surface. 
In fact, if the localization $w_{\bf k_{\|}}$ were not included into 
the calculations or if it were taken closer to the surface the H-{}induced 
peaks would be hidden in the background noise of other electron-{}hole 
excitations. 
The clear image in Figs.~\ref{FSuscMo} and \ref{FSuscW} only appears 
if we take the position of the classical turning point 
into account. 

To our knowledge these are the first theoretical studies of the 
rare-{}gas atom scattering spectrum using a first-{}principles electronic 
bandstructure. 
Our findings support the earlier suggestion~\cite{kohl95a}
that the giant indentations, seen only in HAS, 
are due to the excitation of electron-{}hole pairs.
In HREELS such excitations are less likely, and thus 
the strong anomaly remains invisible. 

\section{Summary and Outlook}\label{SSumm}
In conclusion, the major results of our work are as follows: 
We demonstrate that for both W\,(110) and Mo\,(110) 
the $(1\times 1)$ surface geometry is stable upon hydrogen adsorption. 
There is no evidence for a pronounced H-{}induced 
top-{}layer-{}shift reconstruction, 
a result confirmed by a recent LEED analysis~\cite{arno96a,arno96b}. 
Also, we give a consistent explanation for the H-{}induced 
anomalies in the HAS spectra of W\,(110) and Mo\,(110). 
The H-{}adsorption induces surface states of $(d_{3z^2-r^2},d_{xz})$ 
character that show pronounced Fermi-{}surface nesting. 
The modestly deep anomaly is identified as a Kohn 
anomaly due to those nesting features~\cite{kohl95a,rugg95,kohl96b}. 
For an understanding of the huge dip we stress 
that the scattering of rare-{}gas atoms is crucially influenced 
by interactions with substrate surface {\em wavefunctions }
at the classical turning point~\cite{pete96}. 
Assuming that the He atom couples efficiently to the H-{}induced surface 
states on H/W\,(110) and H/Mo\,(110) we
therefore conclude that the deep HAS anomaly   
is predominantly caused by a direct excitation of
electron-{}hole pairs during the scattering process. 

We hope that our calculations and interpretations stimulate additional 
theoretical and experimental work: 
For instance, it is highly important to better understand the details 
of rare-{}gas atom scattering processes at ``real'' surfaces.
Also, the liquid-{}like behavior of the H-{}adatoms 
observed in the HREELS measurements~\cite{bald94b} 
remains an unresolved puzzle. 
One would also like to know whether the HAS and HREELS spectra of 
the Mo$_{0.95}$Re$_{0.05}$\,(110) alloy system reveal H-{}induced 
anomalies as expected.  
Finally, we call for a new ARP study of the clean and H-{}covered 
Mo\,(110) and W\,(110) surfaces in order to experimentally 
identify the Fermi surface. 
We also note that scattering experiments with atoms or molecules like 
Ne or H$_2$ might provide additional insight into the interesting behavior 
of these surfaces. 

\acknowledgements
We have profitted considerably from discussions with Erio Tosatti. 
In particular, they have enlighted our understanding of phason and 
amplitudon modes as a possible explanation of the two indentations. 

\appendix
\section*{Test Calculations}
\subsection{Bulk}\label{ABulk}
\mediumtext
\begin{table}
\caption{Theoretically and experimentally obtained crystal parameters for 
Mo and W. 
The percentage deviations from experiment are given in parentheses.
The results of this work are calculated non-{}relativistically (N)
and scalar-{}relativistically (R). 
Also, the different XC potentials used in the theoretical 
studies are given: 
Wigner~\protect{\cite{wign34}} (W37), 
Barth and Hedin~\protect{\cite{bart72}} (BH72), 
and  Ceperley and Alder~\protect{\cite{cepe80}} (CA80)
parameterized by Perdew and Zunger 
\protect{\cite{perd81}}. 
The abreviation ``NL-PP'' stands for 
``non-{}local pseudo-{}potential'' method. 
All theoretical numbers ignore the influence of zero-{}point 
vibrations.}
\begin{tabular}{lllll}
metal   &method        &XC
&$a_0$\ [\AA]
&$B_0$\ [MBar]\\
\hline
Mo
&experiment\tablenotemark[1]
        &&3.148                 &2.608\\
&NL-PP\tablenotemark[2]
        &BH72   &3.152 (+0.1)           &3.05 (15.5)\\
&LAPW\tablenotemark[3]
        &BH72   &3.131 ($-$0.5)         &2.91 (11.6)\\ 
\cline{2-5}
&this work (N)          &CA80   &3.126 ($-$0.7)         &2.88 (10.4)\\
&this work (R)          &CA80   &3.115 ($-$1.1)         &2.89 (10.8)\\
\hline
W
&experiment\tablenotemark[4]
        &&3.163                         &3.23\\  
&NL-PP\tablenotemark[2]
        &BH72   &3.173 (+0.3)           &3.45 (6.8)\\
&LAPW\tablenotemark[5]
        &W34    &3.149 ($-$0.4)         &3.46 (7.1)\\
&LAPW\tablenotemark[3]
        &BH72   &3.162 ($\pm$0.0)       &3.40 (5.3)\\ 
\cline{2-5}
&this work (N)          &CA80   &3.194 ($+1.0$)         &2.92 (--9.6)\\
&this work (R)          &CA80   &3.137 ($-0.8$)         &3.37 (4.3)\\
\end{tabular}
\tablenotetext[1]{from Ref.~\onlinecite{kata79}}
\tablenotetext[2]{from Ref.~\onlinecite{zung79}}
\tablenotetext[3]{from Ref.~\onlinecite{matt86}}
\tablenotetext[4]{from Refs.~\onlinecite{shah71,kitt76}}
\tablenotetext[5]{from Ref.~\onlinecite{jans84}}
\label{Tbulk}
\end{table}
\narrowtext
In Table~\ref{Tbulk} we list the calculated equilibrium lattice 
constants $a_0$ and the bulk moduli 
$B_0\equiv-V \partial^2E^{\rm tot}/\partial^2V$, and 
compare them to published results from experiment and theory. 
Our values were obtained using the LAPW parameter setting presented 
in Table~\ref{TParam}. 
\narrowtext

\subsection{Surfaces}\label{ASurf}
\subsubsection{Test System}
The focus of this paper lies on the Mo(110) and W(110) surfaces. 
In order to test the numerical accuracy of our calculations 
for such systems it is important to go beyond typical tests for 
the crystal bulk. 
For instance, we need to know the error bars for atomic geometries 
obtained via forces as compared to a pure total energy calculation. 
Five- and more-{}layer systems are not suitable and in fact for 
most cases also not necessary for such a detailed study. 
Therefore, questions refering to the size of the vacuum region, 
to the accuracy of the directly calculated LAPW forces, and to the 
appropriate wavefunction cutoff will be answered by studying a 
simple Mo(110) double-{}layer slab system. 
Starting from the LAPW parameter set listed in Table~\ref{TParam} 
we vary one parameter and study its influence on the 
evaluated total energy and the atomic force. 
These quantities are calculated with respect to the 
inter-{}layer distance $z$. 
Again, we note that we are interested not in the physics of this 
double-{}layer system 
but in the accuracy and the convergence of our method. 
Therefore, the calculations are done using just four $\bf k_{\|}$ 
points within the SBZ. 

Employing a least-{}square fit which has the form of a 
Morse potential~\cite{mors29}
\begin{equation}
E^{\rm tot}(z)= D \left[1- {\rm e}^{-\beta (z-d-d_0)}\right]^2
\end{equation}
we calculate the equilibrium relaxation parameter $d$. 
Here, $d_0$ represents (110) layer distance in the bcc crystal. 
Additionally, the vibrational properties are of considerable interest. 
From the Morse parameters $D$ and $\beta$ one extracts the oscillator 
frequency by using the expression 
\begin{equation}
\omega^2= \frac{4D\beta}{M_I}
\end{equation}
where the atomic masses $M_I$ are those of the Mo and W nuclei. 
A corresponding fit is also employed for the directly calculated 
forces normal to the (110) surface:
\begin{eqnarray}\nonumber
F(z)
&=&-\frac{dE^{\rm tot}(z)}{dz}\\
&=&-2\tilde D \tilde \beta 
\left[{\rm e}^{-2\tilde \beta(z-\tilde d-d_0)} 
- {\rm e}^{-\tilde \beta(z-\tilde d-d_0)}\right]
\quad. 
\end{eqnarray}
This enables a quantitative study of the accuracy of the evaluated LAPW 
forces with respect to the total energy $E^{\rm tot}(z)$. 
Within the Tables~\ref{TVacuum}--{}\ref{Tvk} the two alternatives 
are marked by ``F''~(force) and ``E''~(energy). 

We consider a calculation to be converged if the relaxation parameter 
$d$ and the frequency $\omega$ differ by less the 
$\Delta d= 0.25\%d^{\rm conv}\approx 0.005$\,\AA\ and 
$\hbar \Delta \omega = \hbar 5\% \omega^{\rm conv} \approx 2$\,meV from the 
fully converged results $d^{\rm conv}$ and $\omega^{\rm conv}$. 
In the following discussion we focus on these two quantities $d$ and 
$\omega$. 
They provide detailed information about the accuracy of 
structure-{}optimization and frozen-{}phonon calculations performed in 
this paper. 

\subsubsection{Vacuum Size}
\begin{table}
\caption{Mo(110) double-{}layer: Morse parameter $d$ and the energy 
$\hbar \omega$ versus the number of vacuum layers.  
The arrow indicates the parameter choice which yields converged results.}
\begin{tabular}{rcccc}
number
 &\multicolumn{2}{c}{$d$\ [$\% d_0$]} 
 &\multicolumn{2}{c}{$\hbar \omega$ [meV]}\\ \cline{2-3}  \cline{4-5}
of layers   
 &E&F&E&F\\ \hline
2& --1.7& --2.1& 36.5& 35.3\\
3& --1.7& --2.1& 36.7& 35.2\\
$\Rightarrow$
4& --2.0& --2.4& 37.3& 35.6\\
5& --2.1& --2.4& 37.4& 35.7\\
6& --2.0& --2.3& 36.9& 35.4\\
\end{tabular}
\label{TVacuum}
\end{table}
The first set of test calculations serves to determine the appropriate 
size of the vacuum region. 
The results presented in Table~\ref{TVacuum} show that an equivalent of 
four substrate layers ($\hat{=}$ 8.8\,\AA) is sufficient to 
decouple the two substrate surfaces accross the vacuum. 
Both relaxation parameter $d$ and the frequency $\omega$ vary only within 
the given error range upon further increasing the vacuum region. 

\subsubsection{LAPW Parameters}
\begin{table}
\caption{Mo(110) double-layer: 
Variation of the wavefunction cutoff parameter $\frac{\hbar^2}{2m}(K^{\rm wf})^2$.  
The notation is equivalent to Table~\protect{\ref{TVacuum}}.}
\begin{tabular}{rcccc}
$\frac{\hbar^2}{2m}(K^{\rm wf})^2$ &
\multicolumn{2}{c}{$d$\ [$\% d_0$]} &
\multicolumn{2}{c}{$\hbar \omega$ [meV]}\\ \cline{2-3} \cline{4-5}
\multicolumn{1}{c}{[Ry]}  &E&F&E&F\\ \hline
9 & --0.5& --1.0& 35.0& 33.4\\
10& --1.5& --2.1& 36.0& 34.6\\
11& --1.6& --2.2& 36.1& 35.2\\
 $\Rightarrow$
12& --2.0& --2.4& 37.3& 35.6\\
13& --1.9& --2.2& 37.1& 35.5\\
14& --1.9& --2.1& 36.5& 35.4\\
\end{tabular}
\label{Tvc}
\end{table}
Another important parameter is the wavefunction energy cutoff 
$\frac{\hbar^2}{2m}(K^{\rm wf})^2$. 
It has a decisive influence on the computer time. 
With respect to the structure parameter $d$ our results in 
Table~\ref{Tvc} indicate that 
the smallest possible value is $\frac{\hbar^2}{2m}(K^{\rm wf})^2=12\,$Ry. 
However, for $\hbar\omega$ already  a calculation with 
$\frac{\hbar^2}{2m}(K^{\rm wf})^2=10\,$Ry 
provides converged results. 

\begin{table}
\caption{Mo(110) double-layer: 
Variation the potential paremters $\frac{\hbar^2}{2m}(G^{\rm pot})^2$ and 
$l^{\rm pot}$.  
The notation is equivalent to Table~\protect{\ref{TVacuum}}.}
\begin{tabular}{rcccc}
\multicolumn{1}{c}{$\frac{\hbar^2}{2m}(G^{\rm pot})^2$}&
\multicolumn{2}{c}{$d$\ [$\% d_0$]}&
\multicolumn{2}{c}{$\hbar\omega$ [meV]}\\ 
\cline{2-3} \cline{4-5}
\multicolumn{1}{c}{[Ry]} &E&F&E&F\\ 
\hline
64 & --2.0& --2.3& 37.3& 35.6\\
81 & --1.9& --2.3& 36.6& 35.5\\
$\Rightarrow$
100& --2.0& --2.4& 37.3& 35.6\\
121& --1.9& --2.3& 37.0& 35.6\\
144& --1.9& --2.3& 37.0& 35.6\\
\end{tabular}

\ \\
\begin{tabular}{rcccc}
\multicolumn{1}{c}{$l^{\rm pot}$}&
\multicolumn{2}{c}{$d$\ [$\% d_0$]}&
\multicolumn{2}{c}{$\hbar\omega$ [meV]}\\ \cline{2-3} \cline{4-5}
&E&F&E&F\\ \hline
2& --1.9& --1.7& 37.5& 34.6\\
$\Rightarrow$
3& --2.0& --2.4& 37.3& 35.6\\
4& --2.0& --2.5& 37.1& 35.2\\
5& --2.1& --2.6& 37.1& 35.3\\
\end{tabular}
\label{Tpot}
\end{table}
The two potential parameters $\frac{\hbar^2}{2m}(G^{\rm pot})^2$ 
and $l^{\rm pot}$ are of little influence 
on the calculated quantities. 
This can be seen in Table~\ref{Tpot}. 
The values of the relaxation parameter $d$ and the frequency $\omega$ 
remain relatively stable within the parameter range 
64\,Ry $<\frac{\hbar^2}{2m}(G^{\rm pot})^2 < 144\,$Ry.  
Both quantities are more sensitive to changes in the  
$(l,m)$ expansion of the MT potential. 
However, a value of $l^{\rm pot}=3$ seems to be sufficient. 

In general we note that energy (E) and force (F) calculations demonstrate 
a similar convergence behavior. 
However, the magnitude of the relaxation in the force calculations 
is overestimated by about $\Delta d = 0.5\%d^{\rm conv}\approx0.01\,$\AA; 
for the frequencies one finds a deviation of about 
$\hbar\Delta\omega =5\,\%\hbar\omega^{\rm conv}\approx1.8$\,meV 
from the converged total energy values. 
Those differences are probably due to numerical inaccuracies in 
the force calculation.

\subsubsection{$\bf k_{\|}$ Point Set \label{Avk}}
Similar calculations for a W(110) double-{}layer system lead to a 
converged LAPW parameter set which is comparable to the one 
obtained for Mo. 
Only the MT potential parameter $l^{\rm pot}$ has to be chosen 
higher: $l^{\rm pot}=4$. 
slightly different.
The agreement between force and total energy calculations 
($\Delta d= 1\%d^{\rm conv}\approx 0.02$\,\AA~and 
$\Delta \omega = 5\% \omega^{\rm conv}$)
is nearly as good as in the case of Mo.
The following calculation is aimed to 
determine a two-{}dimensional $\bf k_{\|}$-point set which can be 
used for the calculation of the atomic and electronic structure 
of Mo\,(110) and W\,(110) surfaces (see Table~\ref{Tvk}). 
\begin{table}
\caption{W(110) double-layer: 
Accuracy of the calculation in dependence 
of the number of  $\bf k_{\|}$-points for the 
(a) valence electrons and (b) semicore electrons. 
The respective fixed $\bf k_{\|}$-point set is made up by 
one special $\bf k_{\|}$-point in the irreducible part of the 
SBZ~\protect{\cite{monk76}}. 
The arrows indicate the  converged $\bf k_{\|}$-point set used 
for the (110) surface calculations. 
The notation is equivalent to Table~\protect{\ref{TVacuum}}.}
\vspace{5mm}
 
(a) valence electrons
\begin{tabular}{rcccc}
\multicolumn{1}{c}{$\#$}
 &\multicolumn{2}{c}{$d$\ [$\% d_0$]} 
 &\multicolumn{2}{c}{$\hbar\omega$ [meV]}\\ \cline{2-3} \cline{4-5}
$\bf k_{\|}$-points &E&F&E&F\\ 
\hline
36 & --8.5& --9.5& 20.5& 21.1\\
49 & --9.0& --10.1& 23.8& 23.1\\
$\Rightarrow$
64 & --8.4& --9.3& 19.2& 20.3\\
81 & --8.4& --9.4& 18.9& 20.4\\
\end{tabular}\vspace*{5mm}

(b) semicore electrons
\begin{tabular}{rcccc}
\multicolumn{1}{c}{$\#$}
 &\multicolumn{2}{c}{$d$\ [$\% d_0$]} 
 &\multicolumn{2}{c}{$\hbar\omega$ [meV]}\\ \cline{2-3}\cline{4-5}
$\bf k_{\|}$-points &E&F&E&F\\ 
\hline
4 & --1.0& --1.8& 27.4& 26.2\\
$\Rightarrow$
9 & --1.0& --1.9& 27.4& 26.2\\
16& --1.0& --1.9& 27.3& 26.2\\
25& --1.0& --1.9& 27.4& 26.2\\
36& --1.0& --1.9& 27.4& 26.2\\
\end{tabular}
\label{Tvk}
\end{table}
Because of the size of (110) surface slab systems it is important 
to keep the number of $\bf k_{\|}$-points as small as possible. 
For the calculations we use the LAPW parameter set given in 
Table~\ref{TParam} and vary the  $\bf k_{\|}$-point 
set for  either the valence or the semicore electrons. 
These sets always consist of uniform two-{}dimensional point meshes. 
Converged results are obtained by using 
64 $\bf k_{\|}$-points 
for the valence and 9 $\bf k_{\|}$-points for the semicore electrons. 

\subsection{Phonons}\label{APhon}
\subsubsection{LAPW parameter}

Our final series of tests is meant to determine the accuracy of the 
frozen-{}phonon calculation. 
In order to keep the computer time for those elaborate calculations 
within pleasant limits one has to keep the number of $\bf k_{\|}$-points 
and basis functions as small as possible. 
As a sample case we study the $\overline S$ point Rayleigh phonon 
of the H/Mo\,(110) system. 

The calculations are performed as follows: 
We distort the atoms of the first substrate layer by 
$t_{1z}(\overline S)=0.08\,$\AA~ according to the pattern in
Fig.~\ref{FPattern} and determine the total energy and force 
changes $\Delta E^{\rm tot}$ and $\Delta F_{1z}$. 
These quantities have the same convergence behavior as 
the phonon frequencies: 
Therefore, if $\Delta E^{\rm tot}$ and $\Delta F_{1z}$ are 
stable with respect to a 
particular LAPW parameter so are the phonon frequencies. 
Converged results for Mo as well as for W are obtained by 
using the following setting: 
wavefunction plane-{}wave cutoff $\frac{\hbar^2}{2m}(K^{\rm wf})^2=10\,$Ry;
$l^{\rm pot}=3$ for the $(l,m)$ expansion of the MT potential; 
uniform $\bf k_{\|}$-point mesh consisting of 56 points; 
five substrate layers; a vacuum region equivalent to four 
substrate layers; XC potential from Refs.~\onlinecite{cepe80,perd81}. 

\subsubsection{$\bf k_{\|}$ Point Mesh \label{SAppPhon}}
In order to accurately describe the electronic structure and 
hence the electron-{}phonon coupling for the $\overline S$ 
point phonon one has to provide a detailed $\bf k_{\|}$-point set.  
For instance, for the study of the $(2 \times 1)$ reconstructions
of the diamond (111) surface Vanderbilt and Louie 
employ a $\bf k_{\|}$-point mesh which becomes logarithmically 
denser close to the zone boundary~\cite{vand84}.
One cannot rely on the results for the two 
Mo\,(110) and W\,(110) double-{}layer slabs in Section\,\ref{Avk} 
because in these systems nesting effects are totally unimportant. 

\begin{table}
\caption{Frozen-{}phonon distortion for H/Mo(110):
Change of the total energy $\Delta E^{\rm tot}$ and the force 
$\Delta F_{1z}$ versus the number of $\bf k_{\|}$-points. 
All $\bf k_{\|}$-point sets are uniform meshes which include the 
symmetry point $\overline{S'}$ of the backfolded SBZ 
(see Fig.~\protect{\ref{FKohn}}). 
Given are also the calculated values for the clean Mo(110) surface 
(with 56 $\bf k_{\|}$-points).}
\begin{tabular}{lcc}
\# $\bf k_{\|}$-points
        &$\Delta E^{\rm tot}$\ [meV]
                &$\Delta F_{1z}$\ [meV/\AA]\\
\hline
16      &110    &672\\
56      &136    &756\\
80      &144    &797\\
120     &135    &790\\
232     &136    &777\\
\hline
clean   &161    &938\\      
\end{tabular}
\label{TFrozEnerDiff}
\end{table}

We perform tests for the H/Mo(110) system using up to 232 
$\bf k_{\|}$-points in the 
SBZ and find that a uniform $\bf k_{\|}$-point mesh 
of 56 points is sufficient in order to obtain converged results 
(see Table~\ref{TFrozEnerDiff}).
For all those frozen-{}phonon calculations the Monkhorst-{}Pack 
$\bf k_{\|}$-point meshes~\cite{monk76} 
are shifted in order to include the $\overline S'$ point 
of the $(1\times1)$ unit cell (see Fig.~\ref{FKohn}). 
For the $\bf k_{\|}$-point summation we employ a Fermi smearing 
of $k_{\rm B}T_{\rm el}\approx 68$\,meV. 
Only for the 232 $\bf k_{\|}$-point calculation listed in  
Table~\ref{TFrozEnerDiff} a smaller value of 
$k_{\rm B}T_{\rm el}\approx 14$\,meV 
is used. 
However, this parameter is apparently uncritical because the energies are 
obtained for the $k_{\rm B}T_{\rm el}\leftarrow 0$\,K limit as discussed in 
Ref.~\onlinecite{neug92}. 

\subsubsection{Basisfunctions}
Next, we check whether the results are sensitive to an increase in the 
wavefunction cutoff, i.e., whether the chosen value 
$\frac{\hbar^2}{2m}(K^{\rm wf})^2=10\,$Ry 
is large enough. 
\begin{table}
\caption{Frozen-{}phonon distortion for H/Mo(110): 
Change of the total energy $\Delta E^{\rm tot}$ and the force 
$\Delta F_{1z}$ with respect to the wavefunction plane-{}wave cutoff
$\frac{\hbar^2}{2m}(K^{\rm wf})^2$.}
\begin{tabular}{lcc}
$\frac{\hbar^2}{2m}(K^{\rm wf})^2$\ [Ry]
        &$\Delta E^{\rm tot}$\ [meV]
                &$\Delta F_{1z}$\ [meV/\AA]\\
\hline
10      &136    &756\\
12      &137    &767\\
\hline
\end{tabular}
\label{TFrozPhonCutoff}
\end{table}
The data listed in Table~\ref{TFrozPhonCutoff} demonstrates that this is 
indeed the case. 
The results remain stable for the higher cutoff 
$\frac{\hbar^2}{2m}(K^{\rm wf})^2=12\,$Ry. 

\subsubsection{LDA versus GGA}
\begin{table}
\caption{Frozen-{}phonon distortion for H/Mo(110):
Variation of the XC potential: 
LDA \protect{\cite{cepe80,perd81}}, GGA \protect{\cite{perd92a}}.}
\begin{tabular}{lcc}
XC Potential    
        &$\Delta E^{\rm tot}$\ [meV]
                &$\Delta F_{1z}$\ [meV/\AA]\\
\hline
LDA     &136    &756\\
GGA     &141    &764\\
\end{tabular}
\label{TFrozXC}
\end{table}

For some systems involving hydrogen the results of a total energy 
calculation are not reliable if the LDA is used.
In those cases the bonding --- especially between hydrogen 
atoms --- is only described 
accurately if gradient corrections are considered~\cite{hamm94}.  
In Table~\ref{TFrozXC} we compare results obtained with the 
generalized gradient approximation 
(GGA)~\cite{perd92a,perd92b} and those found within the LDA. 
The latter is identical with the 56 $\bf k_{\|}$-point calculation 
in Table~\ref{TFrozEnerDiff}. 
There, the XC potential of Cepereley and Alder~\cite{cepe80} parametrized 
by Perdew and Zunger~\cite{perd81} is used. 
The small energy and force differences of 5\,meV and 8\,meV/\AA, 
respectively,  clearly shows that our adsorbate systems are 
rather insensible to the XC formalism chosen.

\narrowtext

\end{document}